# Automated Generation of Test Models from Semi-Structured Requirements

Jannik Fischbach, Maximilian Junker
Qualicen GmbH, Germany
{firstname.lastname}@qualicen.de

Andreas Vogelsang
TU Berlin, Germany
andreas.vogelsang@tu-berlin.de

Dietmar Freudenstein
Allianz Deutschland AG, Germany
dietmar.freudenstein@allianz.de

*Abstract*—[Context:] Model-based testing is an instrument for automated generation of test cases. It requires identifying requirements in documents, understanding them syntactically and semantically, and then translating them into a test model. One light-weight language for these test models are Cause-Effect-Graphs (CEG) that can be used to derive test cases. [Problem:] The creation of test models is laborious and we lack an automated solution that covers the entire process from requirement detection to test model creation. In addition, the majority of requirements is expressed in natural language (NL), which is hard to translate to test models automatically. [Principal Idea:] We build on the fact that not all NL requirements are equally unstructured. We found that 14% of the lines in requirements documents of our industry partner contain "pseudo-code"-like descriptions of business rules. We apply Machine Learning to identify such semi-structured requirements descriptions and propose a rule-based approach for their translation into CEGs. [Contribution:] We make three contributions: (1) an algorithm for the automatic detection of semi-structured requirements descriptions in documents, (2) an algorithm for the automatic translation of the identified requirements into a CEG and (3) a study demonstrating that our proposed solution leads to 86% time savings for test model creation without loss of quality.

## I. Introduction

Testing is crucial in modern development projects and contributes significantly to the success of IT projects. It ensures both the quality of the developed system and its compliance with customer requirements. For complete test coverage, all specified requirements need to be covered by corresponding test cases. Driven by the increasing development speed in agile IT projects, test cases must be created instantaneously in order to receive immediate feedback on the current status of the system. This requires increasing automation in test case creation. A substantial contribution towards a higher degree of automation is *Model-Based Testing* [1]. The desired system behavior is represented by test models and subsequently converted into test cases automatically. Test models do not only serve as an instrument for the automated generation of test cases but also enable early detection of issues in requirement specifications. Furthermore, they allow teams to transparently discuss the desired system setup. In the past, the generation of test cases from test models has been widely automated [2], while practitioners today encounter a variety of problems when creating and maintaining these models [3].

The creation of test models requires a number of tasks, which are currently mostly performed manually. Test designers need to identify and extract relevant information from requirements specifications mostly written in natural language (NL) [4], [5], fully understand them and finally transform them into a suitable test model. Particularly in large software projects, the extraction of these requirements proves to be challenging because the documents quickly become complex and contain a range of other information in addition to the actual requirements (parameter definitions, context information, examples, etc.) [6]. Furthermore, the complexity of the requirements increases simultaneously with growing complexity of IT systems requiring a number of different inputs and outputs to be considered when creating test models. This results in a time-consuming and error-prone test model creation process because the quality of the test model strongly depends on the capabilities of the test designer. The lack of automation of the process also leads to additional maintenance costs. Any change in a requirement requires a revision of the test model.

The notation of requirements significantly influences the creation of a test model. Informal and unstructured descriptions are usually imprecise and ambiguous representing a barrier for the automated transfer into test models. In some domains, such as embedded software, we see an increasing trend towards more structured and model-based specifications of requirements. In other domains, such as business information systems, natural language is still the dominant way to specify requirements. We build on the fact that not all NL requirements are equally unstructured. In fact, in many NL requirements specifications, we see descriptions that clearly exhibits a certain structure. Examples include user stories created along one of the proposed templates (*As an <actor>, I want <goal>, so that <benefit>.*), requirements created along the lines of a constrained natural language (CNL) notation such as EARS [7], or textual descriptions of business rules. Sometimes, even logical constructs such as *If, Then, Else* are embodied in natural language constructs as a way of describing the expected system behavior more precisely [8]. While analyzing 11 requirements documents from our industry partner, an average of 14% of the lines in the documents contained "pseudo-code"-like descriptions of business rules. These requirement descriptions represent a good starting point for the automation of test model creation due to their structured nature. In the remainder of the paper, we will call these parts of requirements specifications *semi-structured requirements*—requirements expressed in a structured form of natural language text. Ideally, we can identify and interpret these requirements within the documents and then convert them into a test model. Drawing on the described problems and our observations gathered in practice, we conclude two main research challenges (RC):

**RC 1:** We need to automatically detect and extract *semi-structured requirements* within requirement documents.

**RC 2:** We need to automatically transform *semi-structured requirements* into test models.

In this paper, we address both research challenges and present an approach that supports practitioners in creating test models. Our approach consists of two steps: (1) automatically identify *semi-structured requirements* in requirements specifications and (2) automatically transform the identified *semi-structured requirements* into *Cause-Effect-Graphs* (CEG), which are well suited to derive test cases according to our preliminary study [9]. An automated transfer of requirements into a CEG would thus complement the test case creation process. Our work makes the following key contributions (C):

**C 1:** We present a Machine Learning based approach for the detection of *semi-structured requirements*. Our proposed method is evaluated on a dataset of 11 requirement specifications provided by Allianz Deutschland.

**C 2:** We introduce a method that allows us to analyze the identified *semi-structured requirements* both syntactically and semantically and to convert them into CEGs.

**C 3:** We present a case study evaluating our algorithm in practice. It leads to a 86% time saving compared to the manual creation without loss of quality in the model.

## II. BACKGROUND AND RELATED WORK

In the following, we describe fundamentals and related work in the field of extracting test models from requirement specifications.

### A. Notation of Requirements

Requirements can be specified in different styles, which can be divided in three different classes: informal, semi-formal, and formal techniques [10], [11], [12]. System behaviour can be described in a narrative form by using unrestricted natural language (informal technique). Informal descriptions do not follow a special syntax or semantics. Semi-formal requirements are expressed in a predefined and fixed syntax that does not have a formally defined syntax. Example of semi-formal requirements descriptions include unrestricted UML models or controlled natural language [7]. Formal notations follow a precise syntax and semantics. An example of such a specification language represents Z, which is based on mathematical notations.

We focus on descriptions that are somewhere between informal and semi-formal descriptions and call them *semi-structured requirements*. *Semi-structured requirements* are similar to controlled natural language. Using control structures such as *When, If, Then, While and Where*, requirements are fed into a specific structure to prevent common issues such as vagueness. An example is the Unwanted Behaviour Pattern in EARS [7], which enforces the formulation as follows: IF <unwanted condition or event>, THEN the <system name> shall <system response>. Hence, EARS requirements follow a clear syntax and the values between the control structures consist of a limited vocabulary. Although control structures are used, our considered requirements do not fit seamlessly into the three described notations categories due to their lack of rigid syntax and limited vocabulary. We therefore consider them to be between informal and semi-formal requirements.

Fig. 1: Exemplary requirements document containing *semi-structured requirements* (see orange dotted frame). This is not an original document of Allianz Deutschland.

### B. Cause-Effect-Graphing

Cause-Effect-Graphing originates from the idea of Elmendorf and Myers [13], [14]. It aims at illustrating the desired system behavior in the form of input and output combinations. The inputs represent the causes and the outputs the respective effects. A *Cause-Effect-Graph* can thus be interpreted as a combinatorial logic network, which describes the interaction of causes and effects by Boolean logic. Both causes and effects are associated by four different relationships: conjunction $\wedge$, disjunction $\vee$, negation $\neg$ and implication $\Rightarrow$. Let $G = (E, C)$ be the CEG shown in Figure 2 with effect set $E$ and cause set $C$. In the example, $|C| = 5$ including $c1$, $c2$, $c3$, $c4$ and

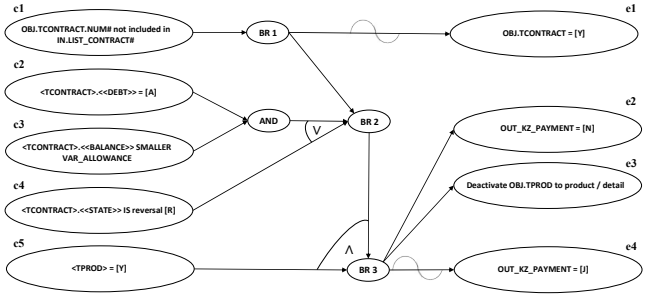

Fig. 2: Exemplary *Cause-Effect-Graph* for Fig. 1.

$c5$ while $|E| = 4$ with $e1$, $e2$, $e3$ and $e4$. Examining the requirement specification in Fig. 1, it emerges that several *If* clauses are nested within each other. This is illustrated by inserting intermediate nodes for each *If* clause resp. "business rule" (BR). To derive test cases from the CEG, the graph is traversed back from the effects to the causes and test cases are created according to specific decision rules, which can be found in [13]. We integrated these rules into the Open Source tool *Specmate*[1] and demonstrated that the tool creates a minimum number of test cases instead of the $2^n$ possible test cases (n is the number of causes) [9]. Thus, CEG models are well suited for balancing between sufficient test coverage and the lowest possible number of test cases. This is also indicated by Table I, which contains the test cases generated by *Specmate* for $G$.

TABLE I: Generated test cases from Fig. 2

| Test Case | Input | | | | | Output | | | |
|---|---|---|---|---|---|---|---|---|---|
| | c1 | c2 | c3 | c4 | c5 | e1 | e2 | e3 | e4 |
| TC 1 | 1 | 0 | 0 | 1 | 0 | 0 | 0 | 0 | 1 |
| TC 2 | 1 | 0 | 0 | 1 | 1 | 0 | 1 | 1 | 0 |
| TC 3 | 0 | 0 | 0 | 1 | 1 | 1 | 0 | 0 | 1 |
| TC 4 | 1 | 1 | 1 | 0 | 1 | 0 | 1 | 1 | 0 |
| TC 5 | 1 | 0 | 1 | 0 | 1 | 0 | 0 | 0 | 1 |
| TC 6 | 1 | 1 | 0 | 0 | 1 | 0 | 0 | 0 | 1 |

*C. State-of-the-Art*

To our knowledge, there is no work that covers the entire process from identifying requirements in documents to the actual translation into test models. A first promising approach for the recognition of pseudocode similar content is provided by Tuarob et al. [15]. The developed method is aimed at the recognition of mathematical algorithms and is therefore not suitable for the detection of requirements. We follow the idea to use Machine Learning for the recognition of pseudo code and design new features which target the characteristics of *semi-structured requirements*.

The translation of functional requirements into test models is an active area of research. There is a number of papers dealing with the translation into State Diagrams [16], [17], [18], Use Case Diagrams [19] and other types of UML Diagrams. So far there is no approach to automatically translate requirements into CEGs. Though Mogyorodi outlines how to manually generate the CEG, there is no automation of the process [20]. We address this research gap and provide a suitable method.

### III. APPROACH: OVERVIEW

Our approach aims at two goals:
1) The recognition of *semi-structured requirements* in requirement documents.
2) The translation of *semi-structured requirements* into CEGs.

For extracting specific content from documents, rule-based methods are usually applied. However, these are not suitable for achieving our first objective. It requires a disproportionately large set of rules to cover different pseudo code spellings (other logical constructs, languages, etc.). The maintenance of these rules is time-consuming, error-prone and causes the recognition to be inflexible. In practice, writing conventions are usually not respected. Throughout the analysis of the requirement documents we observed that often parts of the pseudo code are missing (e.g. the delimiters). Rule-based approaches tend to be too rigid in this respect and therefore lead to wrong results. Our observations coincide with the findings of Tuarob et al. who discovered a similar problem in mathematical pseudo code [15]. We therefore use Machine Learning to meet Goal 1. Instead of focusing on the exact spelling of the pseudo code, we differentiate its general syntax from that of natural language. We address Goal 2 by identifying cause-effect-patterns within the identified requirements. First, we translate the requirement into a syntax tree in order to synthesize the effects and causes. Subsequently, we traverse the tree using a specific visitor pattern, analyze its semantics and translate it into a CEG.

### IV. AUTOMATIC DETECTION OF SEMI-STRUCTURED REQUIREMENTS

We argue that a requirements specification represents a collection of coherent lines that contain either natural language or pseudo code (PC). The detection of PC can be therefore understood as a binary problem classifying lines as follows:

$C_0$: Line does not contain any pseudo code or
$C_1$: Line contains pseudo code

In order to solve this classification problem we implement a algorithm, which assigns the lines to one of the two categories. It first splits the specification into individual lines, then classifies each line based on four different criteria and lastly returns all lines classified as PC. Our detection algorithm therefore implements function $f : \mathbb{R}^4 \to \{C_0, C_1\}$, where each vector represents a single line. The criteria originate from an analysis of requirement specifications provided by Allianz Deutschland and represent the following properties of PC:
1) *Number of special characters:* Whereas in natural language sentences only dots, commas or colons are used to structure the content, pseudo code contains a variety of other special characters. These are usually applied for declaring variables or performing logical queries.
2) *Number of words:* Observing the pseudo code, it becomes evident that the lines are often significantly shorter than

[1] https://specmate.io/

natural language lines. This can be demonstrated by the logical constructs like *If* and *Then*, which are presented in separate lines for a better readability.
3) *Degree of indentation:* In contrast to natural language, pseudo code often does not start at the beginning of a line, but is rather positioned in the middle of a line due to indentations. To cover this property, we count the number of whitespaces between the beginning of the line and the first letter per line.
4) *Capital letters to letters ratio:* A further distinctive characteristic of pseudo code is the intensive use of capital letters to denote variables and logical constructs. To illustrate this property, we calculate the ratio between capital letters used and the total number of letters for each line.

### A. Dataset

Since no scientific work has yet been carried out on the recognition of *semi-structured requirements* on the basis of ML, there is no data set available for this purpose. Hence, we have created our own data set representing 11 requirements specifications from our industry partner. For this purpose, we used a Python script that splits the original documents into single lines, calculates the corresponding feature values for each line, and generates a .csv file. Each line was assigned a number and manually labeled by the first author with either 0 or 1 to indicate pseudo code lines as well as natural language lines to enable *Supervised Learning*. To ensure the validity of the labeling, the second author randomly inspected the data. Our generated data set contains all 11 requirement documents and consists of 4450 lines of which 613 are pseudo code.

### B. Balancing the Training Data

The generated data set was imbalanced. Only 13.8% are positive samples (i.e., lines with pseudo code). To avoid the class imbalance problem, we apply different balancing techniques and compare their impact on the performance of the classifiers. We applied the following methods:
1) Sampling-based Methods: *Random Over Sampling* (ROS), *Random Under Sampling* (RUS), *Tomek Links* (TL), *Synthetic Minority Oversampling Technique* (SMOTE).
2) Cost-sensitive Methods: *Class Weighting* (CL).

### C. Classification Algorithms

We benchmarked 6 different algorithms, which are widely employed for binary classification in research and practice: *Decision Tree*, *Logistic Regression*, *Support Vector Machines*, *Random Forest*, *Naive Bayes* and *K-Nearest Neighbor*.

### D. Evaluation Strategy and Metrics

We follow the idea of *Cross Validation* and divide the data set in two segments: a training and a validation set. While the training set is used for fitting the algorithm, the latter is utilized for its evaluation based on real world data. We opt for a 10-fold *Cross Validation* since a number of studies have indicated that a model that has been trained this way demonstrates low bias and variance [21]. Selecting a metric requires to consider which misclassification (*False Negative* (FN) or *False Positive* (FP)) matters most resp. causes the highest costs. Every pseudo code line has to be identified in order to create the CEG entirely. The costs of FN are therefore significantly higher than those of FP because, otherwise, parts of the CEG are missing and the generated test case is incorrect. As a result, in the tradeoff between *Precision* and *Recall* we clearly opt for the latter.

TABLE II: Evaluation of the classifiers. *Class Weighting* (CL) is not supported by *K-Nearest Neighbor* and *Naive Bayes*.

| Classifier | Balancing | Recall | Precision |
|---|---|---|---|
| Random Forest | ROS | 0.834 | **0.710** |
|  | RUS | **0.910** | 0.628 |
|  | SMOTE | 0.876 | 0.597 |
|  | TL | 0.834 | 0.621 |
|  | CL | 0.824 | 0.608 |
| Logistic Regression | ROS | 0.674 | 0.592 |
|  | RUS | 0.636 | 0.572 |
|  | SMOTE | 0.688 | 0.583 |
|  | TL | 0.152 | 0.392 |
|  | CL | 0.671 | 0.579 |
| Support Vector Machines | ROS | 0.621 | 0.539 |
|  | RUS | 0.613 | 0.567 |
|  | SMOTE | 0.607 | 0.563 |
|  | TL | 0.149 | 0.291 |
|  | CL | 0.622 | 0.559 |
| K-Nearest Neighbor | ROS | 0.829 | 0.631 |
|  | RUS | **0.880** | 0.532 |
|  | SMOTE | 0.841 | 0.642 |
|  | TL | 0.758 | 0.687 |
|  | CL | - | - |
| Decision Tree | ROS | 0.812 | 0.681 |
|  | RUS | 0.837 | 0.617 |
|  | SMOTE | 0.794 | 0.657 |
|  | TL | 0.784 | 0.672 |
|  | CL | 0.795 | 0.674 |
| Naive Bayes | ROS | 0.684 | 0.577 |
|  | RUS | 0.726 | 0.554 |
|  | SMOTE | 0.703 | 0.593 |
|  | TL | 0.257 | 0.548 |
|  | CL | - | - |

### E. Results

Table II presents the performance of the individual classifiers. Analyzing the metrics reveals that *Random Forest* and *K-Nearest Neighbor* are particularly suitable for our use case. They demonstrate the highest *Recall* values and are therefore capable of recognizing of about 90% of the pseudo code lines. Here, however, the choice of the respective balancing method is of crucial importance, as can be observed in the example of *K-Nearest Neighbor*. Applying *Tomek Links* leads to a *Recall* value of 75%, whereas the use of *Random Under Sampling* results in a *Recall* value of 88%. Similarly, *Random Forest* achieves very different values when using certain balancing methods. Considering the respective *Precision* values, it is striking that the algorithms produce a high number of FP. For instance, the *K-Nearest Neighbor* algorithm recognizes 88% of the pseudo code lines, but strongly ignores the *Precision* if the training set has been balanced using *Random Under Sampling*. It classifies too many lines as pseudo code, of which only 53% are actually pseudo code. Ultimately, this leads to an unclean output containing almost all the pseudo code, but also many other lines that are not needed to create the CEG models. This requires that excess lines are manually removed before the CEG can be created automatically. As the *Random Forest* exhibits a greater precision value and consequently a better explanatory power, we select it as a suitable algorithm. The remaining algorithms perform worse than the *Random Forest* with regard to the main decision criterion and are therefore not suitable for capturing pseudo code comprehensively, excluding them from further consideration.

## V. Translation of Semi-Structured Requirements into Test Models

After detecting the pseudo code, the next step is to convert it into a corresponding CEG. This requires the determination of cause-effect-patterns within the pseudo code (syntactic analysis) and the subsequent comprehension of the relationships between the identified causes and effects (semantic analysis). In the following, we outline how we implemented these two analysis steps.

### A. Syntactic Analysis

When synthesizing the cause-effect-patterns, it must be noted that the pseudo code does not follow a rigid grammar. The formulation of the causes and effects conforms to no particular form of notation such as <variable name> <operator> <value>. Our approach is to identify the position of the respective causes and effects based on the logical constructs. We apply a loosely defined grammar[2], which we implemented using ANTLR, and create an *Abstract Syntax Tree* (AST) for each pseudo code (see Fig. 3). It provides an overview of both the inputs to be processed by the system (causes) and the expected system behavior (effects).

### B. Semantic Analysis - Terminology

To reflect the relation between causes and effects in the CEG, we semantically analyze the AST. We traverse the tree from bottom to top (i.e. post-order), create corresponding CEG nodes and link them according to predefined decision rules. For a precise description of these rules we introduce the following terminology.

1) *Parse-Tree-View:* Within the AST, we distinguish between two different types of nodes.

   **Assignment Nodes** *Assignment-nodes* are labeled with "cause" or "effect" (see green nodes in Fig. 3). They do not contain information about the cause or effect itself but serve as a link between them. Assignment nodes have three children, of whom the middle child indicates the type of connection. If this child contains the value "AND", it specifies a conjunction and the node can be considered as an *Assignment-node$_{con}$*. If the value equals "OR", the connection is a disjunction, meaning that the node represents an *Assignment-node$_{dis}$*.
   **Content Nodes** Like assignment-nodes, *Content-nodes* are also labeled as "cause" or "effect"(see orange nodes in Fig. 3). In contrast to the *Assignment-nodes*, their children contain information about the content of causes or effects. The children of *Content-nodes* are leaves in the parse tree. We differentiate between *Content-node$_{cause}$* and *Content-node$_{effect}$*.
   **Business Rule** We define each *If, Then, Else* clause as a *Business-rule*. Nested clauses are therefore indicated by nodes with the label "businessRule" (see blue nodes in Fig. 3).

2) *CEG-Model-View:* The CEG model consists of the following elements.

   **CEG Nodes** *CEG-nodes* represent *Content-nodes* and *Business-rules*. We use the following notation to refer to the *CEG-nodes*: *CEG-node$_{<type>}$* with $type \in \{effect, cause, BR\}$.
   **CEG Connection** All connection types presented in Section II are modeled by *CEG-connections*. We use the following syntax: *CEG-connection$_{<type>}$* with $type \in \{implic, neg, dis, con\}$.

### C. Semantic Analysis - Procedure

The transformation of the AST into a CEG comprises the following three steps. We welcome fellow researchers to inspect, reuse and adapt our source code on Github[2].

*C.1 Translate Content-nodes and Business-rules to CEG-nodes*

In the first step, we create a *CEG-node* for each identified *Content-node* and *Business-rule*. In the example of Fig. 3, one *CEG-nodes$_{cause}$* needs to be created to cover the causes of both the first and third business rule, and three *CEG-nodes$_{cause}$* for the second business rule. Since the pseudo code contains four different effects, another four separate *CEG-nodes$_{effect}$* must be created. Additionally, three *CEG-nodes$_{BR}$* are created.

*C.2 Relate Causes to Business Rules*

For each business rule, the respective causes have to be linked to each other. Four cases can be separated here:

1) The business rule covers only one cause: Create a *CEG-connection$_{implic}$* and link the cause to the corresponding business rule.
2) The causes are connected only by *Assignment-nodes$_{con}$*: Create a *CEG-connection$_{con}$* and associate all causes directly with the corresponding business rule.
3) The causes are connected only by *Assignment-nodes$_{dis}$*: Create a *CEG-connection$_{dis}$* and associate all causes directly with the corresponding business rule.
4) The causes are connected by a combination of *Assignment-node$_{con}$* and *Assignment-node$_{dis}$*: To detect related and isolated causes, iterate over the parse tree and seek the longest sequence of *Assignment-nodes$_{con}$* that are not interrupted by *Assignment-nodes$_{dis}$*. Insert an intermediate node for the *Assignment-nodes$_{con}$* on the lowest depth and connect related causes to this node by a *CEG-connection$_{con}$* to ensure the minimum number of intermediate nodes. Afterward, connect the intermediate node and isolated causes to the business rule using *CEG-connection$_{dis}$*.

---
[2]The algorithm as well as the applied grammar can be found on https://github.com/fischJan/Automated_CEG_Creation.git

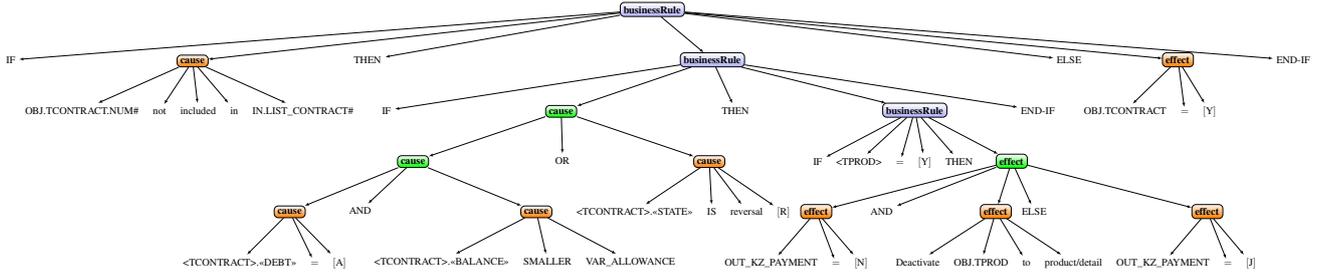

Fig. 3: *Abstract Syntax Tree* representing the requirement specification in Figure 1. *Assignment-nodes* are marked in green, *Content-nodes* in orange and *Business-rules* in blue.

### C.3 Relate Business Rules to Effects

Each effect needs to be associated with its respective business rule. This process is more complex than the previous step, as some exceptions have to be covered due to the nesting of the business rules and their *Else* blocks. In the example of Fig. 3, three business rules are nested within each other. For the effects of a business rule to occur, the causes of the superior business rule need to emerge. A business rule is therefore always a condition for all effects of the subordinate business rules. Here, both the first and second business rule is required so that the effects of the third business rule can arise. In general, a *CEG-node$_{BR}$* must be linked to every effect that is part of a business rule with a higher depth. The degree of depth thus reflects the degree of nesting. Therefore, the algorithm distinguishes between two categories of effects: On the one hand its individual effects and on the other hand the effects of subordinate BRs. Joining the latter it uses *CEG-connections$_{implic}$*, while connecting to individual effects two cases have to be considered:

1) Causes that are not located in the *Else* block are linked to the business rule by *CEG-connections$_{implic}$*.
2) Causes that are located in the *Else* block are linked to the business rule by *CEG-connections$_{neg}$*.

## VI. CASE STUDY

Our study aims to examine whether our approach can speed up the process of CEG creation in practice. We want to answer the following research question: How much time can be saved by using the algorithm compared to a manual CEG creation?

### A. Study Setup

We conducted the case study with 3 test designers, who are experienced in the creation of CEGs. We use three additional documents provided by our industry partner and complete the procedure of CEG creation with each participant. We ask each study participant to translate the pseudo code sections of each document into a CEG first manually (stage ❶) and then by applying the algorithm (stage ❷). The manual creation proceeds as follows: ❶a search for the pseudo code section, ❶b translate the pseudo code into a CEG. The application of the algorithm includes the following sub-steps: ❷a execute the detection algorithm (automated), ❷b compare the output with the pseudo code in the document including additions/deletions (manual) and ❷c execute the translation algorithm (automated). The difference between both required time efforts illustrates the impact of the algorithm on the CEG generation process. Here, it is important that the three documents differ in their complexity. The first document (201 lines) contains only a single short pseudo code section (5 lines). The second document (327 lines) includes 2 pseudo code sections which are significantly longer (11 and 17 lines). The third document (437 lines) shows the greatest complexity as it consists of two nested pseudo code sections, which are nested multiple times (19 and 23 lines). This difference in complexity permits a particularly detailed insight into usability and reveals potential weaknesses in handling complex documents.

TABLE III: CEG creation times in the case study (in min.)

|               | # | Manual | Tool-Supported | Time Saving |
|---------------|---|--------|----------------|-------------|
| Participant 1 | 1 | 02:04  | 00:33          | 73.4%       |
|               | 2 | 11:31  | 01:07          | 90.3%       |
|               | 3 | 16:21  | 02:44          | 83.3%       |
| Participant 2 | 1 | 01:54  | 00:36          | 68.4%       |
|               | 2 | 10:05  | 01:20          | 86.8%       |
|               | 3 | 13:40  | 01:18          | 90.5%       |
| Participant 3 | 1 | 01:21  | 00:35          | 56.8%       |
|               | 2 | 05:36  | 00:48          | 85.7%       |
|               | 3 | 08:42  | 00:57          | 89.1%       |
| Mean          |   | 07:55  | 01:06          | 86.1%       |

### B. Study Results

The use of our semi-automated approach results in an average time saving of 86% (see Table III). Although the process has not yet been fully automated, it is already leading to a significant improvement in the day-to-day business of test designers. The manual effort required to compare the identified pseudo code lines with the actual existing ones was within reasonable time limits. All participants consider the significant time saving as the main reason to apply the algorithm in practice. A comparison of the manually created CEGs with the CEGs created by our approach reveals that their content is the same. Hence, the quality of the created test models does not suffer due to the use of the algorithms and is consequently suitable as a replacement for manual creation. The algorithm is particularly suitable for handling complex requirement documents. Multiple nested pseudo code sections contain a number of different causes and effects as well as logical constructs. As a result, they are difficult to read and therefore opaque, which may lead to mistakes during manual creation. According to all study participants, this represents the largest field of application for the algorithm, as it not only helps recognizing the pseudo code but also to interpret it. This has been particularly evident in the translation of the third test document. Initially, all participants had to understand the structure of the pseudo code before they could translate it. During this process some causes and effects were overlooked, which required several adjustments with negative effects on the creation time. We hypothesize that this became considerably easier by using the algorithm and that our approach leads to less errors when processing requirements documents.

## VII. LIMITATIONS AND THREATS TO VALIDITY

The detection of *semi-structured requirements* defined in requirements documents could not be fully automated. The majority of the pseudo code lines were detected during the evaluation, though approx. 10% had to be inserted manually afterwards. Due to the low *Precision* value, additional adjustments are required to remove the lines that are not relevant for CEG creation. Regarding the translation into the CEG, our approach depends on a consistent syntax of the pseudo code. In case of spelling and grammar errors, the translation fails. Future research should employ compensation techniques such as calculating the *Levensthein Distance* to make the translation more robust. A threat to internal validity is social pressure in the execution of group experiments, especially when recording times. To prevent the test designers from competing with each other in the creation of the CEG, we inspected the three test documents individually with the study participants. Furthermore, none of the study participants was informed of the time spent by the others. The translation of *semi-structured requirements* into a CEG is subject to potential maturation effects. However we argue that this effect can be neglected since the actual creation of the CEG (step 2c) was performed automatically and very little of the second phase of the study was performed manually (step 2b). With regard to external validity, it can be stated that some parts of our approach need to be adapted for use across other domains. This requires extending the detection algorithm with additional features and small modifications of the ANTLR grammar to handle further notation styles.

## VIII. CONCLUSIONS AND OUTLOOK

Test models have proven to successfully describe the desired system behavior and derive suitable test cases. They therefore constitute an integral part of modern development projects, but their creation and maintenance is very time-consuming, error-prone and is usually performed manually. We present an approach that generates *Cause-Effect-Graphs* from *semi-structured requirements*. Specifically, we extract relevant requirements from requirements documents using Machine Learning and analyze them both syntactically and semantically to create the corresponding CEG. Our study with 3 experts on 3 exemplary cases shows that the process could not yet be fully automated but the application of our algorithm results in 86% time savings. In future work, we see potential to transfer our approach to user stories and their acceptance criteria. Specifically, we want to translate acceptance criteria expressed in the Gherkin format into CEGs and thus tailor the test model generation even more to the agile context.